\documentstyle[12pt]{article}

\newcommand{\mincir}{\raise -2.truept\hbox{\rlap{\hbox{$\sim$}}\raise5.truept
\hbox{$<$}\ }}
\newcommand{\magcir}{\raise -2.truept\hbox{\rlap{\hbox{$\sim$}}\raise5.truept
\hbox{$>$}\ }}
\newcommand{\siml}{\raise -2.truept\hbox{\rlap{\hbox{$\sim$}}\raise5.truept
\hbox{$<$}\ }}
\newcommand{\simg}{\raise -2.truept\hbox{\rlap{\hbox{$\sim$}}\raise5.truept
\hbox{$>$}\ }}

\newcommand{\be}{\begin{equation}}
\newcommand{\ee}{\end{equation}}
\newcommand{\ba}{\begin{eqnarray}}
\newcommand{\ea}{\end{eqnarray}}
\newcommand{\brr}{\begin{array}}
\newcommand{\err}{\end{array}}
\newcommand{\bc}{\begin{center}}
\newcommand{\ec}{\end{center}}

\newcommand{\hm}{\,h^{-1}{\rm Mpc}}
\newcommand{\vel}{\,{\rm km\,s^{-1}}}

\begin{document}
\baselineskip 15pt

\begin{titlepage}

  \begin{flushright}
    TUM-HEP-288/97
    \\
    SFB-375/206
    \\
   August 1997
  \end{flushright}

\begin{center}
  {\Large \bf The Formation of Cosmic Structures in a Light Gravitino
    Dominated Universe}
\vskip 1cm
{\large
    Elena Pierpaoli$^{a)}$, Stefano Borgani,$^{b)}$  Antonio Masiero$^{a,c)}$
    \\
and Masahiro Yamaguchi$^{d)}$\footnote{On leave of absence from Department of 
Physics, Tohoku University, Sendai 980-77, Japan.}}

\vskip 0.4cm 
{\it a) SISSA-International School for Advanced Studies \\
       via Beirut 2-4, I-34013 Trieste, Italy} \\

{\it b) INFN, Sezione di Perugia \\
    Dipartimento di Fisica, Universit\`{a} di Perugia \\
    via A. Pascoli, I-06100 Perugia, Italy} \\

{\it c) Dipartimento di Fisica,  Universit\`{a} di Perugia and  \\
   INFN, Sezione di Perugia, via A. Pascoli, I-06100 Perugia, Italy} \\

{\it d) Institute f\"{u}r Theoretische Physik \\
   Physik Department, Technische Universit\"{a}t M\"{u}nchen \\
   D-85747 Garching, Germany}

\vskip 0.5in

\abstract $~$ We analyse the formation of cosmic structures in models
where the dark matter is dominated by light gravitinos with mass of $
100$ eV -- $1$ keV, as predicted by gauge--mediated supersymmetry (SUSY)
 breaking models.  After evaluating the number of degrees of freedom at the
gravitinos decoupling ($g_*$), we compute the transfer function for
matter fluctuations and show that gravitinos behave like warm dark
matter (WDM) with free-streaming scale comparable to the galaxy mass
scale.  We consider different low--density variants of the WDM model,
both with and without cosmological constant, and compare the
predictions on the abundances of neutral hydrogen within
high--redshift damped Ly--$\alpha$ systems and on the number density
of local galaxy clusters with the corresponding observational
constraints. We find that none of the models satisfies both
constraints at the same time, unless a rather small $\Omega_0$ value
($\mincir 0.4$) and a rather large Hubble parameter ($\magcir 0.9$) is
assumed. Furthermore, in a model with warm + hot dark matter, with hot
component provided by massive neutrinos, the strong suppression of
fluctuation on scales of $\sim 1\hm$ precludes the formation of
high--redshift objects, when the low--$z$ cluster abundance is
required. 
 We conclude that all different  variants of a light gravitino DM
dominated model show strong difficulties for what concerns 
cosmic structure formation.
 This gives a severe cosmological constraint on the gauge-mediated SUSY
   breaking scheme.

\end{center}
\end{titlepage}

\baselineskip 0.7cm

\section{Introduction}

Since the moment (early 80's) that low-energy supersymmetry (SUSY) was
invoked in gauge unified schemes to tackle the gauge hierarchy problem
\cite{Nilles}, it became apparent that it had also a major impact on
several cosmological issues. By far the most studied consequence was
the presence of a stable SUSY particle in all models where a discrete
symmetry, known as R-parity, is imposed to prevent the occurrence of
baryon and lepton renormalizable terms in the superpotential. Indeed,
R-parity assigns a different quantum number to ordinary particles and
their SUSY partners. Hence the lightest SUSY particle (LSP) is
absolutely stable and constitutes, together with photons and
neutrinos, a viable candidate for relic particles of the early
Universe.

The two best candidates we have to play the role of LSP are: the
lightest neutralino (i.e. the lightest among the fermionic partners of
the neutral gauge and Higgs fields) and the gravitino (the fermionic
partner of the graviton in the gravity multiplet)
\cite{SUSYDM-review}. Which of the two is the actual LSP strictly
depends on the mechanism one envisages for the SUSY breaking, or, more
precisely, for the transmission of the breaking of SUSY from some
hidden sector to the observable sector of the theory (ordinary
particles and their superpartners belong to this latter sector). If
the ``messengers" of the SUSY breaking are of gravitational nature (as
it happens in the more ``orthodox'' supergravity models), then the
lightest neutralino is likely to be the LSP. In these schemes the
gravitino mass sets the scale of SUSY breaking in the observable
sector and, hence, it is expected to be in the $10^2 - 10^3$ GeV
range. On the other hand, it has been vigorously emphasized recently
(after ten years of silence about this alternative) that gauge,
instead of gravitational interactions may be the vehicle for the
transmission of the SUSY breaking information to the observable sector
\cite{Dineetal}. In these scenarios the scale of SUSY breaking is much
lower than in the supergravity case and consequently, as we will see
below, the gravitino mass is much lower than $10^2$ GeV.  Hence in
this class of gauge mediated SUSY breaking (GMSB) models the gravitino
is more likely to play the role of LSP with a mass which can range a
lot, depending on the specific scale of SUSY breaking, say from a
fraction of eV up to O(GeV).

>From a cosmological point of view, the neutralino LSP scenario with a
lightest neutralino in the tens of GeV range constitutes an ideal
ground for a cold dark matter (CDM) proposal \cite{cdm}. Indeed there
exists a  sufficiently vast area of the SUSY parameter space where such an
LSP becomes non relativistic at a sufficiently early epochs so as to
make its free--streaming mass much smaller than the typical galaxy
mass scale ($\sim 10^{11}M_\odot$). The standard version of the CDM
scenario, with $\Omega_0=1$ for the density parameter, $h=0.5$ for the
Hubble parameter\footnote{We take $H_0=100 \,h \vel$ Mpc$^{-1}$ for
  the Hubble constant.} and $P(k)\propto k$ for the post--inflationary
power spectrum of Gaussian adiabatic density fluctuations, is
generally accepted to fail in reproducing several observational tests.
On scales of few tens of $\hm$ it develops a wrong shape of the power
spectrum \cite{PeacDodds}. Furthermore, once normalized to match the
detected level of CMB temperature anisotropies \cite{cobe}, it produces
too large fluctuations on scales $\mincir 10\hm$, with a subsequent
overproduction of galaxy clusters \cite{clabu}.

These failure of the standard CDM model may be overcome in these SUSY
models by finding a way to suppress fluctuations on $10\hm$ scales,
without decreasing too much power on the $\sim 1\hm$ scale, which
would delay too much the galaxy formation epoch. A first possibility
is adding to the LSP CDM candidate some massive light neutrino
(Cold+Hot DM model; CHDM) to provide about 20--30\% of the critical
density \cite{primack96}. This has just the effect of decreasing the
fluctuation amplitude around the neutrino free--streaming scale, so as
to change the power--spectrum in the right direction. A further
possibility is assuming a density parameter substantially smaller than
unity, either with or without a cosmological constant to provide
spatial flatness \cite{lowom}. A lower cosmic density gives rise to a
larger horizon size at the matter--radiation equality epoch, so as to
increase the large-to-small scale power ratio in the spectrum of
cosmic density fluctuations.

If the gravitino is the LSP, one loses the traditional CDM candidate,
being such a gravitino a more likely warm (WDM) candidate, its
free--streaming mass scale being comparable to the galaxy mass scale
\cite{BMY,KSY}. This happens when its mass lies in the range [0.1--1]
keV, which represents the situation that we will analyse in detail in
this paper.

It is already known that just replacing the cold LSP with a warm one
in the standard CDM scenario does not provide a viable scenario for
the formation of cosmic structures \cite{colombi}. Indeed, the effect
of introducing the warm component is that of suppressing fluctuations
only at the galaxy mass scale, while leaving the power spectrum
unaffected on the cluster mass scales, where standard CDM fails.

Therefore, if we desire a GMSB scheme to provide the dominant DM
content of the Universe, we need some prescription to improve the WDM
scenario. To this purpose, we will analyse in the following what
happens if we follow the same pattern as for improving CDM, namely
either adding a hot neutrino component or lowering the density parameter.
Our analysis will focus on the interesting class of GMSB schemes,
although many of our conclusions may equally well apply to models with
a generic WDM other than the light gravitino.

The purpose of our analysis is twofold. On one hand, given the success
of suitable CHDM and low--density CDM models in accounting for several
observational constraints (in particular providing a low level of
density fluctuations at the $10 h{^-1}$Mpc scale to avoid cluster
overproduction, while having enough power at about $1 h^{-1}$Mpc to
form galaxies at an early enough epoch), we ask whether the agreement
can be kept when a warm gravitino component replaces the cold
candidate.  On the other hand, from a more particle physics oriented
point of view, we would like to make use of the cosmological
constraints related to the DM issue to infer constraints on the GMSB
models, in particular shedding some light on the range of the allowed
(or at least cosmologically favoured) scales of SUSY breaking in this
class of theories.

The paper is organized as follows. In Section 2 we present the general
features of the GMSB models, focusing in particular on their predicted
light gravitinos. We compare the two scenarios, gravity-mediated and
gauge-mediated SUSY breaking, in relation to their LSP predictions and
implications for DM. We provide the main tools for the computation of
the relic gravitino abundance in GMSB models. Section 3 describes the
scenarios for the formation of cosmic structures when the DM content
is dominated by light gravitinos.  Here we compute the corresponding
power spectra of density fluctuations at the outset of recombination.
Afterwards, we present the observational data that we will use to
constrain this class of models, namely the abundance of neutral
hydrogen within high-redshift damped Ly--$\alpha$ systems and the
number density of local galaxy clusters.  In Section 4 we compare the
model predictions for the formation of cosmic structures with the
abovementioned data. The main conclusions of our analysis are
summarized in the final Section 5.

\section{Light gravitinos in SUSY}

In a supersymmetric model \cite{Nilles}, each ordinary particle is
associated with a superpartner.  We assign R-parity even to the
ordinary particles and odd to their superpartners.  In supergravity,
that is a natural extension of the supersymmetric standard models to
the framework of local supersymmetry, we have another R-odd particle,
the gravitino, which is the superpartner of the graviton.  The
lightest of the R-odd particles, namely the lightest superparticle
(LSP), is absolutely stable, under the assumption of the R-parity
conservation, which was originally introduced in order to avoid too
fast proton decays. The LSP is thus a dark matter candidate, if its
expected relic abundances lie within a suitable range of values.

As a starting point, we review some properties of the gravitino.
Imposing the vanishing cosmological constant in the Einstein
supergravity Lagrangian, one finds that the gravitino mass is related
to the SUSY breaking scale $\Lambda_{SUSY}$  as follows:
\begin{equation}
     m_{\tilde G}=\frac{1}{\sqrt{3}}\frac{\Lambda_{SUSY}^2}{M_{Pl}},
\end{equation}
where $M_{Pl}$ is the reduced Planck mass $\sim 2.4 \times 10^{18}$
GeV. On the other hand, the soft SUSY breaking masses for the
superparticles are given as
\begin{equation}
m_{soft} \sim \frac{\Lambda_{SUSY}^2}{M},
\end{equation}
where $M$ effectively represents the mass scale of the interactions
that transmit the breakdown of SUSY in the hidden sector to the
observable sector, the latter including particles of the SUSY standard
model.  We call $M$ the messenger mass scale.  In the conventional
scenario of the gravity-mediated SUSY breaking, the transmission is
due to gravitational interactions. In this case, the messenger mass
scale is $M \sim M_{Pl}$, so that the gravitino mass will be
comparable to the other soft masses.  In order to have the soft masses
at the electro-weak scale the SUSY breaking scale should be at an
intermediate scale $\sim \sqrt{m_W M_{Pl}}$.

On the other hand, one can consider the case where the SUSY breaking
is transmitted by gauge interaction. The idea of the gauge mediation
\cite{gauge-80s} is older than the gravity-mediation, and has recently
been revived with fruitful results \cite{Dineetal}.  In this
case the gauge interaction can set the messenger mass scale much lower
than the Planck mass. Since the soft masses are fixed at the
electro-weak scale, the SUSY breaking scale can be much smaller than
the intermediate scale of $\sqrt{m_W M_{Pl}}$. Correspondingly the
gravitino can be much lighter than the other superparticles.  Now a
crucial question is: how light is the gravitino?  The answer should
depend on the details of the messenger of the SUSY breaking.  In most
of the gauge-mediated models, there are three independent sectors.
They are the hidden sector, the messenger sector and the observable
sector.  The interaction between the last two sectors is the
standard-model gauge interaction, so its strength is fixed.  But the
interaction between the first two is model dependent, so is the
messenger mass scale.  For example, in the original models of
gauge-mediation \cite{Dineetal} it was shown \cite{deGMM} that
$\Lambda_{SUSY}$ cannot be smaller than $10^7 $ GeV, the corresponding
gravitino mass being $\sim 10^2$ keV.  However, a lighter gravitino
should be possible from viewpoints of both model building and
phenomenology.  In the SUSY gauge-mediated approach the soft masses
arise at the loop level (to avoid the supertrace constraint
\cite{supertrace}). Hence a lower bound on $\Lambda_{SUSY}$ is provided by
the relation
\begin{equation}
    \Lambda_{SUSY} \simg \frac{16 \pi^2}{g^2} m_{soft}
\end{equation}
where $g$ is some gauge coupling constant.  For instance, recently Izawa et al.
\cite{INTY} have constructed a model where $ m_{soft} \siml 0.1 g^2/16\pi^2
\Lambda_{SUSY} $.  In this case, $\Lambda_{SUSY}$ can be as small as $O(10^5)$
GeV for $m_{soft}=O(10^2)$ GeV.  In view of the above consideration, in this
paper we consider the following gravitino mass range \footnote{In the framework
  of no-scale models, one may consider a somewhat larger range of the gravitino
  masses \cite{no-scale}.}
\begin{equation}
   1 \ \mbox{eV} \ \siml m_{\tilde G} \siml \mbox{a few TeV}
\end{equation}

It is noteworthy that interaction of the longitudinal component (spin
1/2 component) of the gravitino is fixed by the low-energy theorem.
Namely the would-be Goldstino has a derivative coupling to the
supercurrent with $1/ \Lambda_{SUSY}^2=1/\sqrt{3} m_{\tilde G} M_{Pl}$
suppression. After integration by parts and the use of equations of
motion \footnote{Here we present the formulas for massless gauge
bosons. One needs to modify the formulas when the gauge bosons get
massive due to symmetry breakdown. In our numerical computation in
Section 2.2 this correction is taken into account.}, we obtain the
following effective Lagrangian \cite{Fayet},
\begin{eqnarray}
  {\cal L}_{eff}= \frac{m_{\lambda}}{8 \sqrt{6} m_{\tilde G} M_{Pl}}
                  \bar {\tilde G}
            [\gamma_{\mu},\gamma_{\nu}] \lambda F_{\mu \nu}
        + \frac{m_{\chi}^2-m_{\phi}^2}{\sqrt{3}m_{\tilde G}M_{Pl}} 
          \bar{\tilde G} \chi_L
             \phi^* +h.c.,
\end{eqnarray}
where $\tilde G$ represents the longitudinal component of the
gravitino (the Goldstino) and $m_{\lambda}$, $m_{\chi}$ and $m_{\phi}$
are the masses of a gaugino $ \lambda$, a chiral fermion $\chi$ and
its superpartner $\phi$, respectively.  The point is that as the
gravitino mass gets smaller the interaction becomes stronger. What
happens physically is that a lighter gravitino corresponds to a
lighter messenger scale, and therefore the Goldstino which is in the
hidden sector has a stronger interaction to the fields in the
observable sector.  This point is crucial when we discuss the
cosmology of the light gravitino.

\subsection{Two Scenarios: Neutralino LSP and Gravitino LSP}  

Among the superparticles which appear in the supersymmetric standard
models, a neutralino tends to be the lightest one and, therefore, it
is stable. The neutralino LSP with mass of the order of 100 GeV turns
out to be a good candidate for the cold dark matter (CDM)
\cite{SUSYDM-review}.  In gravity-mediated models with $m_{\tilde G}
\sim 10^2$--$10^3$ GeV, we face the traditional gravitino cosmological
problems \cite{gravitino-problem}.  Namely, unless gravitinos are
strongly diluted at inflation and they are not regenerated in the
reheating phase ($T_{reh} \siml 10^8$ GeV), they would spoil the
canonical picture of big-bang nucleosynthesis (BBN).

On the other hand, if the gravitino is lighter than the neutralino,
the latter is no longer stable, and decays to the gravitino.  It was
pointed out \cite{MMY} that its decays would also destroy the BBN if
its life time is sufficiently large. A limit on the life time depends
on the abundances of the neutralinos before decay. We quote here a
conservative bound of $10^6$ sec as an upper bound for the life time of
the neutralino from the BBN constraint.

In this case  the gravitino will be the stable
LSP. Suppose that the spin 1/2 components of gravitinos were in
thermal equilibrium at an early epoch.\footnote{We assume that the
Universe underwent inflationary era, so that the spin 3/2 components
of the gravitinos were not thermalized after that.} As temperature
went down, the processes which kept the gravitinos in equilibrium
became ineffective and they decoupled from the thermal bath. After
that, the number of gravitinos per comoving volume was frozen out.
This freeze-out took place while the gravitinos were relativistic.
Following a standard procedure \cite{KolbTurner}, one can calculate
the relic density of the gravitinos \cite{PagelsPrimack}
\begin{eqnarray}
\label{eq:omgr}
     \Omega_{\tilde G} h^2 &= & 0.282 \mbox{eV}^{-1} m_{\tilde G} Y_{\infty}
\nonumber \\
      & = & 1.17 \left( \frac{100}{g_*} \right) 
                 \left( \frac{m_{\tilde G}}{10^3 \mbox{eV}} \right),
\end{eqnarray}
where $\Omega_{\tilde G}$ is the contribution of the (thermal)
gravitinos to the density parameter, $h$ is the Hubble parameter in
units of 100 km/s/Mpc, and $g_*$ stands for the effective degrees of
freedom of relativistic particles when the freeze-out of the
gravitinos takes place. Note that $g_*=106.75$ for the full set of
particle contents of the minimal standard model and $g_*=228.75$ for those
of the minimal supersymmetric standard model.  Thus one expects that
$g_*$ at the freeze-out will fall somewhere in between the two
numbers. The computation of $g_*$ is a crucial point of our analysis
and we will come back to it later on. For later convenience, we
introduce the yield, $Y_{\infty}$, of the gravitinos, defined by
\begin{equation}
   Y_{\infty}=\left( \frac{n_{\tilde G}}{s} \right)_{\infty}
             =\frac{0.617}{g_*}, \label{yield}
\end{equation}
where $n_{\tilde G}$ is the number density of the gravitinos and $s$
is the entropy density.  The subscript $\infty$ means that the ratio
is evaluated at a sufficiently late time (i.e., low temperature) at
which it is constant.

We will first briefly discuss the case when the relic abundance of the
gravitinos calculated in this way exceeds the closure limit,
$\Omega_{\tilde G} \simg 1$.  This corresponds to the gravitino mass
region $m_{\tilde G} \simg 1$ keV $(g_*/100) h^2$. In this case, as
was discussed in Refs.~\cite{MMY,deGMM}, entropy production is needed
to dilute the gravitino abundance in order not to overclose the
Universe. To avoid an excessive reproduction of the gravitinos after
the entropy production, its reheating temperature must be low; its
upper bound varies from $10^3$ to $10^8$ GeV, depending on the
gravitino mass.  The lower the gravitino mass is, the lower the
reheating temperature should be.  If the reheating temperature happens
to saturate the upper bound quoted above, the gravitinos will dominate
the energy density of the Universe, and play the role of DM.

On the other hand, the low reheating temperature required by the
closure limit leads to the question of how to generate the baryon
asymmetry of the Universe.  Since the reheating temperature can be
still higher than the electro-weak scale, baryogenesis during the
electro-weak phase transition may work for some region of the
parameter space \cite{EWbaryogenesis}.  Another possibility is to use
the Affleck-Dine mechanism, which was explored in detail in
Ref.~\cite{deGMM} in the framework of the gauge-mediated SUSY
breaking.

When the gravitino mass is smaller than $(g_*/100) h^2$ keV, the
thermal relic density of the gravitinos $\Omega_{\tilde G}$ is smaller
than one.  This is the region that we will study in detail in this
paper. As we discussed previously, models providing this range for the
gravitino mass can be devised. It is also interesting to mention that
a possible explanation of the $ee\gamma \gamma$ event \cite{CDFevent}
at CDF by the light gravitino scenario \cite{gravitino-interpretation}
requires this range of gravitino mass; otherwise the neutralino would
not decay into a photon and a gravitino inside the detector.  A
particularly interesting parameter region for cosmology is the region
in which $0.1 \siml \Omega_{\tilde G} \siml 1$ is realized, and thus
the gravitino mass density constitutes a significant portion of the
density of the whole Universe. A DM particle with mass within   the
sub-keV to keV range is known as warm dark matter \cite{wdm,BMY,KSY}.
Differently from CDM, it is characterized by having a sizable free
streaming length until matter-radiation equality, roughly of the order
of Mpc, but still much smaller than that of the hot dark matter (HDM),
like a few eV neutrino.  We will discuss scenarios of cosmic structure
formation within a WDM dominated Universe in the following sections.

If, instead, the gravitino mass is as small as to give $\Omega_{\tilde
G} \ll 0.1$, then it becomes cosmologically irrelevant and an
alternative DM candidate is required.

\subsection{Computation of $g_*$}

Before moving to the discussion of cosmic structure formation, we would
like to come back to the question of $g_*$, the effective degree of
freedom of relativistic particles at the freeze-out of gravitinos.  Of
particular interest is the region where $m_{\tilde G} \siml 1$ keV so
that gravitinos of thermal origin dominate the energy density of the
Universe. The crucial relevance of $g_*$ lies in the fact that, for a
specified value of $\Omega_{\tilde G} h^{2}$, it fixes the
corresponding gravitino mass and, therefore, the free--streaming
scale.

The production and destruction rates of the gravitinos due to
scattering processes are proportional to the fifth power of the
temperature and, therefore, their abundance rapidly drops down as the
temperature decreases.  Thus, decay and inverse decay processes are
more important for a light gravitino whose freeze-out occurs at a
rather low temperature \cite{MMY}. In the following we will focus on
these processes.

The relevant Boltzmann equation can be casted in the form
\begin{equation}
     \dot{n}_{\tilde G} +3 H n_{\tilde G} =C,
\end{equation}
where $n_{\tilde G}$ is the gravitino number density and $H$ is the
expansion rate of the Universe.   As a collision term, we consider
contributions from two body decay (and inverse decay) processes
\begin{equation}
      C= \sum_{a,b} \Gamma (a\rightarrow b \tilde G)
         \left\langle \frac{m_a}{E_a} \right \rangle
         n_a \left ( 1- \frac{n_{\tilde G}}{n_{\tilde G}^{eq}}
\right).
\end{equation}
Here $\Gamma (a \rightarrow b \tilde G) $ is the partial width of the
species $a$ to $b$ and $\tilde G$, $\langle m_a/ E_a \rangle$ stands
for the thermal average of the Lorentz boost factor, with $m_a$ and
$E_a$ being mass and energy of $a$, $n_a$ is its number density and
finally the superscript ``$eq$'' indicates the equilibrium value of a
given quantity. After some algebra, the above Boltzmann equation can
be rewritten as
\begin{eqnarray}
& &   Y' -\frac{s'}{3s} R Y \,= \,-\frac{s'}{3s} R Y^{eq},
\label{Boltzmann} \\
& &   R \, = \, \frac{\sum \Gamma (a \rightarrow b \tilde G) 
              \langle m_a/E_a \rangle n_a/n_{\tilde G}^{eq}}{H},
\label{R}
\end{eqnarray}
where $Y$ is the yield of the gravitinos as defined by
Eq.~(\ref{yield}), and the apex symbol denotes derivative with respect
to the temperature.  Eq.~(\ref{Boltzmann}) can be solved to give
\begin{equation}
   Y(T) =Y^{eq}(T) +\int^{T_0}_T dT' 
        \exp\left( -\int^{T'}_{T} dT'' R(T'')s'/3s \right) 
        Y^{eq \ \prime} (T').
\label{sol-Y}
\end{equation}
Here the temperature $T_0$ is taken to be sufficiently high so that 
the gravitino is still in thermal equilibrium. 

In order to understand the meaning of Eq.~(\ref{sol-Y}), let us
consider the case where $R(T)$ changes abruptly at a temperature $T_f$
such as $R(T)=\infty$ for $T> T_f$ and $0$ for $T < T_f$.  In this
case we can approximate $\exp( -\int^{T'}_{T} dT'' R(T'')s'/3s)$ with
a step function $\theta (T_f-T')$, so that
\begin{equation}
   Y(T)=Y^{eq}(T) + \int^{T_0}_T dT' \theta(T_f-T') Y^{eq \ \prime}(T')
       =Y^{eq} (T_f),
\end{equation}
thus reproducing the usual result.  In the present case, however,
$R(T)$ gradually decreases as a species becomes non-relativistic.
Therefore, we need to integrate Eq.~(\ref{sol-Y}) numerically to
evaluate $Y(T)$ accurately.  For sufficiently low $T$, the yield
$Y(T)$ approaches its constant value $Y_{\infty}$, from which we
obtain $g_*$ using Eq.~(\ref{yield}).

Results are presented in Table 1.  We show the value $g_*$ for a range
of model parameters.  In this computation, we assumed a typical sparticle
mass spectrum in a simple class of gauge-mediated models
\cite{DGP}. Explicitly, we take for the gauginos
\begin{equation}
     M_{1}=\frac{5}{3} \frac{\alpha_1}{4 \pi} \Lambda_G, \
     M_{2}= \frac{\alpha_2}{4 \pi} \Lambda_G, \ 
     M_{3}= \frac{\alpha_3}{4 \pi} \Lambda_G,
\end{equation}
and for the sfermion masses
\begin{equation}
    m^2= 2 \left[ C_3 \left( \frac{\alpha_3}{ 4\pi} \right)^2
              + C_2 \left( \frac{\alpha_2}{ 4\pi} \right)^2 
              +\frac{5}{3} \left(\frac{Y}{2} \right)^2
               \left( \frac{\alpha_1}{ 4\pi} \right)^2 \right] \Lambda_S^2.
\end{equation}
In the above expressions $\alpha_i$ is a gauge coupling constant in the standard model,
$Y$ is a hypercharge of $U_Y(1)$, while $C_3=4/3$ for a $SU(3)_C$
triplet, $C_2=3/4$ for a $SU(2)_L$ doublet, and $0$
otherwise. $\Lambda_G$, $\Lambda_S$ are introduced to parameterize the
transmission of SUSY breaking from the messenger sector to the
observable sector.\footnote{To avoid further complication, we set a
light Higgs mass to be the $Z^0$ mass, and masses of heavier Higgs and
higgsinos to be the same as the left-handed slepton mass. Furthermore
we did not include D- or F-term contributions to the scalar
masses. Also we ignored the mixing in the mass matrix of the
neutralino and the chargino sector.} We provide $g_*$ values for two
cases: (a) the right-handed slepton mass $m_{\tilde l_R}$ equals to the
bino mass $M_1$, {\it i.e.} $m_{\tilde l_R}=M_1$, and 
(b) $m _{\tilde l_R}= 2 M_1$. In both cases, we find
that $g_*$ is around 100 for a wide range of the parameter space.
For a given $\Omega_{\tilde G}$, a lower value of $g_*$ implies a lighter
gravitino, making structure formations at small scales more difficult,
as we will discuss in the following sections.  The fact that $g_*$
tends to lie in the lower side should be kept in mind, though we will
explore a somewhat wider range for $g_*$.

\begin{table}
\begin{tabular}{|lr|rrrrrr|}
 \hline
\multicolumn{2}{|c|}{ (a)}        & 
\multicolumn{6}{l|}{ $m_{\tilde G}$ (eV)} \\ 
\multicolumn{2}{|c|}{$m_{\tilde l_R}=M_1$}
   & 10 & 50 & 100 & 200 & 500 & 1000 \\
\hline
$M_1$ (GeV) 
      & 50 & 87 & 93 & 101 &110 & 122 & 136 \\ 
      & 100 & 87 & 89 & 93 & 111 &114 & 124 \\
      & 150 & 87 & 89 & 92 & 97 & 109 & 119 \\
      & 200 & 88 & 90 & 93 & 97 & 105 & 115 \\
\hline
\end{tabular}

\bigskip

\begin{tabular}{|lr|rrrrrr|}
 \hline
\multicolumn{2}{|c|}{ (b)}       & 
\multicolumn{6}{l|}{ $m_{\tilde G}$ (eV)} \\ 
\multicolumn{2}{|c|}{$m_{\tilde l_R}= 2 M_1$}
 & 10 & 50 & 100 & 200 & 500 & 1000 \\
\hline
$M_1$ (GeV) 
      & 50  & 87 & 91 & 95 & 102 & 116 & 128 \\
      & 100 & 87 & 90 & 93 & 98 & 107 & 116  \\
      & 150 & 88 & 91 & 93 & 97 & 104 & 111  \\
      & 200 & 88 & 92 & 94 & 98 & 103 & 108  \\
\hline
\end{tabular}
\end{table}

Table 1: Value of effective degrees of freedom of relativistic particles,
at the gravitino freeze-out, $g_*$, as a function of the gravitino mass
$m_{\tilde G}$ and the $U(1)_Y$ gaugino mass $M_1$.  In the case (a) the
right-handed slepton mass is $m_{\tilde l_R}=M_1$, and in (b) $m_{\tilde
  l_R}=2M_1$.

\section{Light gravitinos and cosmic structure formation}

\subsection{Computation of the transfer functions}
The fundamental quantity that allows to make predictions about the
formation of cosmological structures, once the underlying Friedmann
background is fixed, is the transfer function $T(k)$, which convey all
the informations about the evolution of a density fluctuation mode at
the wavenumber $k$ through the matter--radiation equality and
recombinations epochs. In the following we will discuss how the
transfer functions for the models under consideration are computed.  As
for models containing only the warm gravitinos (WDM) we will consider
the $\Omega_0\le 1$ cases, both with ($\Lambda$CDM) and without (OCDM)
a cosmological constant term, $\Omega_\Lambda =1-\Omega_0$, to restore
the spatial flatness. Furthermore, we will consider also the class of
$\Omega_0=1$ mixed models, whose DM content consists both of warm
gravitinos and one species of hot neutrinos, having mass $m_\nu\simeq
91\,\Omega_\nu h^2$ eV ($\Omega_\nu$ is the neutrino contribution
to the density parameter).

Here we will only sketch our implementation of the Boltzmann code to
compute $T(k)$ and we refer to the relevant literature
(\cite{maber95}; \cite{Lidlyth93}) for more technical details.
 
The transfer function is defined as
\be 
T(k)\, =\, {{\sum_{i=1}^{N_s} \Omega_i \delta_{i,z=0}} \over {\sum_{i=1}^{N_s}
\Omega_i \delta_{i,z=z_i}}}\,,
\label{eq:tgen}
\ee
where $N_s$ is the number of different massive species in the model, $\delta_i$
is the energy overdensity of the $i-th$ component and
$z_i$ a suitable initial redshift such that the smallest considered
scale is much larger than the horizon scale at $z_i$.

We evaluate the transfer function for the models of interest in two
steps: firstly we solve the equations for the fluctuation evolution of
all the species involved in the models (namely the baryons, the
radiation, the massless and massive neutrinos, and the gravitinos) for
a number of $k$ values; secondly, we find a suitable analytic
expression which is able to provide a good fitting to the transfer
functions for the whole class of considered models, by varying a
minimal set of parameters.

As for the fluctuation evolution, the goal is to find the final
amplitude $\delta_i$ for the different species, given the initial one.
This goal is achieved in different ways for different components.  For
baryons only two differential equations must be solved: one regarding
their overdensity and one for their velocity; for relativistic
particles it is necessary to solve a hierarchy of coupled differential
equations for the coefficients of the harmonic expansion of the
perturbation in order to well describe the free--streaming behaviour .

For massive free--streaming particles, different free--streaming
behaviours can be expected depending on which fraction of particles
has to be considered still relativistic at a certain epoch.  For
these species it is therefore necessary to follow the fluctuation
evolution separately for particles having different momenta.  A
representative set of different values of the momentum is chosen, and
the density fluctuation evolution is evaluated for each value of this
set.  The overall $\delta_i$ is therefore found by integrating the
zero--th order harmonic coefficients over the momentum, with weights
chosen on the basis of the distribution function.  This is the reason
why, unlike for CDM, for massive free--streaming components the shape
of the spectral distribution function affects the shape of the final
transfer function.
  
In our case, both gravitinos and massive neutrinos
have an initial thermal distribution,
so the equations describing their evolution are qualitatively  the same for 
both the components. 
 What makes the difference between the two is the
redshift at which they become non--relativistic, being higher for the
warm $\tilde G$ than for hot $\nu$. As a consequence, such two
particle populations will be characterized by different
free--streaming scales.

All the calculations were performed in the syncronous gauge.  For a
detailed description on how a thermal free--streaming component is
treated in the syncronous gauge, see Mah and Bertschinger
\cite{maber95}.  From a numerical point of view, we find that a higher
degree of accuracy is needed when dealing with WDM--dominated models
if compared to the CDM--dominated ones.  The reason is that all the
$\delta_i$ are coupled by means of the potential; whose evolution
equation, in turn, depends upon all the the different overdensities,
each of them contributing with a weight $\Omega_i$.  If the
overdensity of the most abundant component is not well evaluated, the
error propagates via the potential to all the other components, and
over time.  In the case of standard MDM, CDM plays this role, it
stabilizes the value of the potential so that a lower accuracy in the
integrals over the momenta of the hot component is allowed.

In the models considered hereafter, gravitinos and massive neutrinos
are the most abundant components, and their overdensities are
evaluated by mean of integrals.  It is therefore necessary to choose
the integration method that, at the same time, $(i)$ provides the best
accuracy, and $(ii)$ minimizes the number of values of the momentum
over which the integration is performed, so as to keep the number of
differential equations to be solved as small as possible.

Within the class of Gauss integration methods \cite{numrec}, we
verified that, keeping fixed the number of integration points in
momentum space, Gauss--Legendre integration performs better than
Gauss--Laguerre, especially for high values of $k$ .  Furthermore, we
found that using Gauss--Legendre integration, 20 integration points
are adequate to obtain stable results.


We computed the transfer function up to $k_{max}\simeq 1$ Mpc$^{-1}$
(for $\Omega_0=1$ and $h=0.5$), with higher $k$ values requiring too
high an accuracy to be reached within a reasonable computational
time. We will show in the following that such a $k_{max}$ value is
larger than the free--streaming wavenumber, $k_{fs}$. Therefore, we
expect that the behaviour of the transfer function at $k> k_{max}$ has
a marginal influence on the hierarchical clustering regime at
$k<k_{fs}$, we are interested in.

In order to provide an analytical fitting to the transfer functions
for the class of purely WDM models, we resorted to the expression
provided by Bardeen et al. \cite{bardeen86}
\be
T_{WDM}(k)\,=\,T_{CDM}(k)\,\exp\left(-{kR_{fs}\over
    2}-{(kR_{fs})^2\over 2}\right) ,
\label{eq:twdm}
\ee
where
\be
T_{CDM}\,=\,{\ln(1+2.34q)\over 2.34q}\,\left[
  1+3.89q+(16.1q)^2+(5.46q)^3+(6.71q)^4\right]^{-1/4}
\label{eq:tcdm}
\ee 
is the transfer function for CDM models. Here, $q=k/\Gamma h$ and
the expression for the shape parameter, $\Gamma=
\Omega_0h\exp(-\Omega_B-\sqrt{2h}\Omega_B/\Omega_0)$ accounts for the
presence of a non--negligible baryon fraction $\Omega_B$
\cite{sugiyama95}.

Therefore, by fitting the transfer function, as computed by the
Boltzmann code, with eqs.(\ref{eq:twdm}) and (\ref{eq:tcdm}) one
obtains the value for the free--streaming scale, $R_{fs}$. More in
detail, our procedure to estimate $R_{fs}$ proceeds as follows.
\begin{description}
\item[(a)] We run the Boltzmann code assuming $\Omega_0=1$ and taking
  $g_*=100$ and 200; the first value is rather representative of
  realistic cases, while the larger $g_*$ corresponds to a very cold
  $\tilde G$ population.
\item[(b)] The free--streaming scale for the $\Omega_0<1$ cases is
  then computed by resorting to the scaling relation $R_{fs}\propto
  m_{\tilde G} \propto \Omega_{\tilde G}$ (cf. eq.(\ref{eq:omgr})),
  where $\Omega_{\tilde G}= \Omega_0-\Omega_B$.
\end{description}

As a result, we find that 
\be
R_{fs}\,=\,0.51\,(\Omega_{\tilde G}h^2)^{-1}\left({g_*\over
    100}\right)^{-4/3} {\rm Mpc}
\label{eq:rfs}
\ee 
always provides an accurate fitting of the exponential suppression of
fluctuations on small scales. We note that our value for $R_{fs}$ is
larger by a factor $\sim 2.5$ than that given by Kawasaki et al.
\cite{KSY}. This difference mainly comes from the fact that our value
is directly obtained by fitting the exactly computed transfer
function, while their value comes from the usual relation between
$R_{fs}$ and $z_{nr}$ (see, e.g., eq.(9.88) in the Kolb \& Turner book
\cite{KolbTurner}), the redshift at which gravitinos becomes non
relativistic, that represents an approximation to the $R_{fs}$ value.
We also confirm the warning by Bardeen et al. \cite{bardeen86}, who
pointed out that the exponential cutoff in eq.(\ref{eq:twdm})
marginally underestimates the transfer function on intermediate
scales, $0.1\mincir k \mincir 0.5 \,(\Omega_0h^2)^{-1}$ Mpc$^{-1}$.
However, we did not attempt here to look for a more accurate fitting
expression, since {\em (a)} the effect is always quite small ($\mincir
5$--10\%) and {\em (b)} we will mainly concentrate our analysis on the
small scales relevant to galaxy and galaxy cluster formation.

We plot in Figure 1 the $T_{WDM}(k)$ shape for $\Omega_0=1$ for
different $g_*$ values (left panel) and for two $\Omega_0<1$ cases
(right panel), also comparing with the corresponding CDM cases. It is
apparent the power suppression on small scales, which depends both on
$g_*$ and on the parameters of the Friedmann background (cf.
eq.(\ref{eq:rfs})).

As for the warm + hot DM (WHDM) case, transfer functions have been
computed for $\Omega_\nu =0.1,0.2,0.3,0.4$ and 0.5 in the case of only
one massive neutrino (cf. ref.\cite{primack96} for the effect of
introducing more than one massive $\nu$), taking $g_*=100$ and 200 and
always assuming $\Omega_0=1$. The analytical fitting is provided by
eq.(\ref{eq:twdm}), where the CDM transfer function is replaced by the
CHDM one, as provided by Pogosyan \& Starobinski \cite{pogstar95}.
Taking $\Omega_{\tilde G}=1-\Omega_\nu-\Omega_B$, we find that
eq.(\ref{eq:rfs}) always provides an accurate fitting to the
exponential cutoff in the transfer function. The shapes of
$T_{WHDM}(k)$ are plotted in Figure 2, showing both the effect of
changing $g_*$ at fixed $\Omega_\nu$ (left panel) and the effect of
changing $\Omega_\nu$ at fixed $g_*$ (right panel). 

According to Figs. 1 and 2, it turns out that the effect of replacing
the CDM component with light gravitinos of mass given by
eq.(\ref{eq:omgr}) is that of eliminating the hierarchical clustering
below some free--streaming mass scale.  In order to provide an
estimate of the free--streaming mass scale, we resort to the almost
Gaussian cutoff at large $k$, to define it as
\be
M_{fs}\,=\,(2\pi R_{fs}^2)^{3/2}\bar \rho \,\simeq \,
0.55\,\left({g_*\over 100}\right)^{-4}(\Omega_{\tilde G}h^2)^{-3} 
\Omega_0 h^2 M_{12}\,,
\label{eq:mfs}
\ee
where $\bar \rho$ is the average cosmic density and
$M_{12}=10^{12}M_\odot$. Therefore, eq.(\ref{eq:mfs}) provides
the limiting mass for the development of hierarchical clustering:
structures of smaller masses form after structure of mass larger than
$M_{fs}$, as a product of their fragmentation. As a consequence, we
expect that a crucial constraint for the whole class of WDM--dominated
models will come from the abundance of high--redshift cosmic
structures.

Having fixed the expression for the transfer function, we define the
power spectrum of the density fluctuations as
$P(k)=AT^2(k)k^{n_{pr}}$, where $n_{pr}$ is the primordial
(post--inflationary) spectral index. The amplitude $A$ is determined
by following the recipe by Bunn \& White \cite{BunnWhite} to normalize
both low--density flat and open models to the 4--year {\sl COBE}
data. In the following, we will not consider the case of
non--negligible contribution  of tensor mode fluctuations to the CMB
anisotropies. Indeed, such an effect would lead to a smaller spectrum
amplitude, with a subsequent delay of the galaxy formation epoch that,
as we will see, represents a major problem for WDM--dominated models.

We plot in Figure 3 the r.m.s. mass fluctuation $\sigma_M$ for the
same models whose $T(k)$ have been plotted in Fig. 1. This quantity is
defined as
\be
\sigma_M^2\,=\,{1\over 2\pi^2}\,\int_0^\infty
dk\,k^2\,P(k)\,W^2(kR_M)\,,
\label{eq:sigmam}
\ee where the length scale associate to the mass scale $M$,
$R_M=(4\pi\bar \rho/3)^{-1}M^{1/3}$, is the radius of the top--hat
sphere whose Fourier representation is given by $W(x)=3(\sin x -x\cos
x)/x^3$. For each model, the corresponding free--streaming mass scale
corresponds to the transition from heavy to light curves in Fig. 3,
while the completely light curves represent the corresponding CDM
cases. It is apparent that such a scale is always at least of the
order of a large galaxy halo. The flattening of $\sigma_M$ at small
masses represents the imprint of non--hierarchical clustering.  On the
other hand, it turns out that the behaviour on the scales of galaxy
clusters, $\sim 10^{15}h^{-1}M_\odot$, is rather similar as for the
CDM--dominated case. In the following we will use the abundance of
local galaxy clusters and of high--redshift protogalaxies, through
data about damped Ly-$\alpha$ systems, to constrain the whole class of
WDM--dominated models. Constraints on larger scales, like bulk--flows
data \cite{vbulk}, are much more similar to the CDM case.

\subsection{Observational constraints}

\subsubsection{High-redshift objects}
The first constraint that we consider comes from the abundance of
neutral hydrogen (HI) contained within damped Ly--$\alpha$ systems
(DLAS; see ref.\cite{Wolfe93} for a review about DLASs). DLAS are
observed as wide absorption through in quasar spectra, due to a high
HI column density ($\magcir 10^{20}$ cm$^{-2}$). Since at $z\magcir 3$
the fractional density of neutral hydrogen associated with DLASs,
$\Omega_{HI}$, is comparable to that associated to visible matter in
local galaxies, it has been argued that DLASs trace a population of
collapsed protogalactic objects. In this context, a crucial question
is to understand whether the observed $\Omega_{HI}$ provides a fair
representation of the collapsed gas fraction at a given
redshift. Effects like gas consumption into stars, amplification
biases due to gravitational lensing of background QSOs
\cite{BartLoeb96} and dust obscuration \cite{FallPei96} could well
alter final results. However, such effects are believed to play a role
at low redshift ($z\sim 1$--2), while they are expected to be less
relevant at the highest redshifts at which DLAS data are available.
For this reason, we will consider as the most constraining datum the
value of $\Omega_{HI}$ reported by Storrie--Lombardi et
al. \cite{Storrie95} at redshift $z\simeq 4.25$ and will assume that
all the HI gas at that redshift is involved in the absorbers.

Several authors recognized DLASs as a powerful test for DM models
using both linear theory and numerical simulations \cite{dlasth}. The
recent availability of high--resolution spectra for several DLAS
systems, allowed Prochaska \& Wolfe \cite{Proch97} to use the internal
kinematics of such systems to severely constrain a CDM model.

In order to connect model predictions to observations, we consider the
fraction of DM which at redshift $z$ is collapsed into structures of
mass $M$,
\be
\Omega_{coll}\,=\,{\rm erfc}\left[{\delta_c(z)\over \sqrt
    2\,\sigma_M(z)}\right]\,.
\label{eq:omcoll}
\ee
Accordingly, $\Omega_{HI}=\Omega_B\Omega_{coll}$. Here, $\sigma_M(z)$
is the r.m.s. fluctuation at the mass scale $M$ at redshift $z$ within
a top--hat sphere. Furthermore, $\delta_c(z)$ is the critical density
contrast whose value predicted by the model for the collapse of a
spherical top--hat fluctuation in a critical density universe, 
$\delta_c=1.69$ independent of the redshift, has been confirmed by
N--body simulations \cite{LaceyCole}. In our analysis we used the
expressions for $\delta_c(z)$ provided in ref.\cite{KitaSuto} for
both low--density flat and open universes. We note, however, that
at the redshift z=4.25, that we are considering, the resulting $\delta_c$
value is always very close to 1.69.

We note that the Press \& Schechter approach \cite{PrSc}, on which eq.
(\ref{eq:omcoll}) is based, holds only in the case of hierarchical
clustering. In our case of WDM models, hierarchical clustering only
takes place on mass scales larger than $M_{fs}$. On smaller scales,
the lack of fluctuations causes the flattening of $\sigma_M$.
Therefore, by estimating $\sigma_M$ at arbitrarily small masses, one
obtains the r.m.s. fluctuations at the free--streaming mass scale. In
our approach, we will give up the dependence on mass scale $M$, which
amounts to assume that DLASs are assumed to be hosted within
protostructures of mass of about $M_{fs}$; protostructures of smaller
mass, instead, are produced later by fragmentation of larger lumps.

As for the observational value of $\Omega_{HI}$, Storrie--Lombardi et
al. \cite{Storrie95} provided for $\Omega_0=1$, $\Omega_{HI}=(1.1\pm
0.2)\times 10^{-3}\,h^{-1}$ at $z=4.25$. In the light of all the above
uncertainties in directly relating $\Omega_{HI}$ to $\Omega_{coll}$,
we prefer to maintain a conservative approach here and to consider a
model as ruled out if it predicts $\Omega_{HI}$ to be less than the
observational 1$\sigma$ lower limit. At this level of comparison we
do not consider as reliable to put constraints to model producing too
high a $\Omega_{HI}$ value.

Furthermore, we should also rescale appropriately the value by
Storrie--Lombardi to include the more general $\Omega_0<1$
cases. Therefore, the limiting value that we consider is
\be
\Omega_{HI} \,= \, 0.0009\,h^{-1}f(\Omega_0,\Omega_\Lambda,z=4.25)\,,
\ee
where
\ba
& &f(\Omega_0,\Omega_\Lambda,z) \,=\, \left({1+\Omega_0z\over
    1+z}\right)^{1/2} ~~~~~~;~~~~~~ \Omega_\Lambda=0 \nonumber \\
& &f(\Omega_0,\Omega_\Lambda,z) \,=\,
{[(1+z)^3\Omega_0+\Omega_\Lambda]^{1/2} \over (1+z)^{3/2}} 
 ~~~~~~;~~~~~~ \Omega_\Lambda=1-\Omega_0\,.
\label{eq:dlas}
\ea

\subsubsection{The cluster abundance}
As for the cluster abundance, it has been recognized to be a sensitive
constraint on the amplitude of the power spectrum 
\cite{clabu}. Based on the Press \& Schechter approach \cite{PrSc},
it is easy to recognize
that the number density of clusters with mass exceeding a given value is
exponentially sensitive to the r.m.s. fluctuation on the cluster mass scale.
Fitting the local X--ray cluster temperature
function with the Press--Schechter approach \cite{PrSc} led several
authors to obtain rather stringent relationships between $\sigma_8$,
the r.m.s. fluctuation value within a top--hat sphere of $8\hm$
radius, and $\Omega_0$ \cite{sig8Om}. In the following we will resort
to the constraint by Viana \& Liddle, who provided the most
conservative and, probably, realistic estimate of errors, mostly
contributed by cosmic variance effects on the local cluster population: 
\ba
& &\sigma_8\Omega_0^{\alpha(\Omega_0)}\,=\,0.60^{+0.22}_{-0.16} \nonumber
\\
& &\alpha(\Omega_0)\,=\,0.36+0.31\Omega_0-0.28\Omega_0^2~~~~;~~~~\Omega_\Lambda=0
\nonumber \\
& &\alpha(\Omega_0)\,=\,0.59-0.16\Omega_0+0.06\Omega_0^2~~~~;~~~~\Omega_\Lambda=
1-\Omega_0\,,
\label{eq:clabu}
\ea
with uncertainties corresponding to the 95\% confidence level.

\section{Discussion}

As for the purely WDM models, we plot in Figure 4 the constraints on
the $(\Omega_0,h)$ plane, for $g_*=150$, from DLAS and cluster
abundance. Only scale--free primordial spectra (i.e., $n_{pr}=1$) are
considered here. Left and right panels correspond to the low--density
flat ($\Lambda$WDM) and open (OWDM) cases, respectively. The solid
line delimiting the coarsely shaded area indicates the limit
for the region of the parameter space which is allowed by the observed
$\Omega_{HI}$ in DLASs: model lying below such curves should be
considered as ruled out, since they produce a too small $\Omega_{HI}$
value at $z=4.25$. The cluster abundance constraint by
eq.(\ref{eq:clabu}) is represented by the finely shaded
region. The dashed curves connect models having the same age of
the Universe: $t_0=9,11,13,15$ and 17 Gyrs from upper to lower curves.

As a main result, we note that there is almost no overlapping between
the regions allowed by the two observational constraints: for fixed
values of the Hubble parameter, cluster abundance tends to select
relatively smaller $\Omega_0$ in order to satisfy the
low--normalization request of eq.(\ref{eq:clabu}). On the other hand,
the DLAS constraint favour higher density parameters, which has the
effect of both decreasing the free-streaming scale and to increase the
small--scale power even in the absence of any free--streaming. Judging
from this plot, one would conclude that the whole class of
gravitino--dominated WDM models would be ruled out by combining
constraints on the cluster and on the galaxy mass scale. A residual
possibility seems to exist to reach a concordance for $\Omega_0\mincir
0.4$ ($\Omega_0\mincir 0.5$) and a high Hubble parameter, $h\magcir 1$
($h\magcir 0.9$) for OWDM ($\Lambda$WDM) models. However, two main
problems arise in this case: {\em (a)} all the current determinations
of the Hubble constant indicates $0.5 < h < 0.8$ \cite{Freed96}; {\em
  (b)} the resulting age of the Universe would be definitely too
small, especially for OWDM models, even on the light of the new
recalibration of globular cluster ages, based on the recent data from
the Hypparcos satellite \cite{globcl}.

We also checked the possibility of considering non--scale-free
primordial spectra ($n_{pr}\ne 1$), although results are not
explicitly presented here. We verified that assuming either blue
($n_{pr}>1$) or red ($n_{pr}<1$) spectra does not improve the
situation. In the first case, power is added on small scales, with the
result that smaller $\Omega_0$ are allowed by DLASs.  However, the
price to be paid is a rapid increase of the cluster abundance, that
also pushes toward smaller $\Omega_0$ the finely shaded area. As for red
spectra, the opposite situation occurs: the reduction of small--scale
power leads both constraints to favour relatively larger $\Omega_0$
values, with no overlapping with the two allowed regions of the
$(\Omega_0,h)$ plane ever attained.

As a matter of fact, the situation becomes even worse when considering
$\Omega_0=1$ WHDM models. Results for this class of models are
reported in Figure 5 on the $(\Omega_\nu,n_{pr})$ plane. Left and
right panels are for $h=0.5$ and 0.6, respectively; smaller $h$ values
are disfavoured by $H_0$ determinations, while larger values are
constrained by the age of the Universe. In both cases the regions allowed
by DLAS and cluster abundance are largely disjoined, especially as
higher $\Omega_\nu$ are considered. Indeed, increasing the neutrino
fraction has the effect of further reducing the power on small scales,
thus further suppressing the high--redshift galaxy formation. 

Based on such results we should conclude that none of the variants of
the WDM gravitino--dominated scenario is able to account at the same
time for the relatively small abundance of clusters at low redshift
and for the relatively high $\Omega_{HI}$ in collapsed structures at
high redshift. It is worth reminding that this result has been
obtained with the rather conservative choice of $g_*=150$. As we have
shown in the previous section, more realistic value of $g_*$ should be
even smaller, thus putting WDM--dominated model in an even worse
shape.

Which are the consequences of such results on the low--energy SUSY
breaking models that we described in Section 2? Of course, a first
possibility is that gravitinos were so light as to be irrelevant from
the point of view of cosmic structure formation. For instance, the
current understanding of high--energy physics phenomenology would
surely allow for $m_{\tilde G}\sim 1$ eV. In this case,
$\Omega_{\tilde G}$ would be negligible. Of course, since $\tilde G$
represents the LSP, the source for a cold DM component should be found
in this case outside the spectrum of SUSY particles 
(e.g., axions).\footnote{See however a recent proposal that a sneutrino 
in the messenger sector can be a CDM candidate\cite{sneutrino}.}

On the other hand, if a scenario with $m_{\tilde G}\sim 100$ eV will
turn out to be preferred, a non--negligible $\Omega_{\tilde G}$ can
not be escaped. In this case, three possible alternative scenarios can
be devised. The first one is to allow for cold + warm DM. However,
since gravitinos have a much smaller free--streaming scale than
neutrinos with $m_\nu\sim 5$ eV, this scenario would suffer from the
same pitfalls of the standard CDM one, unless one takes $\Omega_0 <
1$. The second possibility would be to have a substantially larger
$g_*$, so that gravitinos behave much like CDM. However, as we have
seen in Section 2.2, it is not clear how a substantially larger $g_*$
can be attained within plausible SUSY models.  The third possibility
would be to abandon the assumption of Gaussian fluctuations in favour
of texture seeded galaxy formation \cite{texture}, which would ease
the formation of high redshift objects. However, also this
possibility has been recently shown to suffer from serious troubles in
producing a viable power spectrum of density fluctuations
\cite{notex}, which make texture--based models as virtually ruled out.

One may argue that the gravitino abundances will be diluted to a
cosmologically negligible level by late--time entropy production. On
the other hand, as the low value of $g_*$ suggests, the reheat
temperature after the entropy production should be lower than the
electro-weak scale to avoid the re-thermalization of the gravitinos,
which severely constraints possible ways to generate the baryon
asymmetry of the Universe.

\section{Conclusions}
In this paper we analysed the cosmological consequences of assuming
the dark matter to be dominated by light gravitinos with mass in the
range $\simeq 100$ eV -- 1 keV, as predicted by gauge mediated SUSY
breaking (GMSB) models. We pointed out that gravitinos with such a
mass behave like warm dark matter (WDM), since their free--streaming
mass scale is comparable to the typical galaxy mass scale. 

After estimating the number of degrees of freedom of relativistic
species at the gravitino decoupling, $g_*$, we resorted to a Boltzmann
code to compute the appropriate WDM transfer functions. These are used
as the starting point to compare gravitino--dominated model
predictions to observational data about the abundance of HI within
high--redshift damped Ly--$\alpha$ systems and about the abundance of
local galaxy clusters.

The main results of our analysis can be summarized as follows.

\begin{description}
\item[(a)] Low--density WDM models with both flat ($\Lambda$CDM) and
  open (OCDM) geometry can not satisfy the two observational
  constraints at the same time, unless a rather small $\Omega_0$ value
  ($\mincir 0.4$) and a rather large Hubble parameter ($\magcir 0.9$)
  are assumed. However, such requests would conflict with
  measurements of the Hubble constant and with current constraints
  about the age of the Universe.
\item[(b)] As for warm + hot (WHDM) models, we find that they have an
  even harder time. The combined free--streaming of both neutrinos and
  gravitinos generates a strong suppression of fluctuations at $\sim
  1\hm$ scale. This makes extremely difficult to form high--redshift ($z\sim
  4$) protogalactic objects if we require the model to match the
  low--$z$ cluster abundance.
\end{description}

Based on such results we claim that no variant of a light gravitino DM
dominated model is viable from the point of view of cosmic structure formation.
Therefore, in the framework of GMSB models, this amounts to require the
gravitino to be light enough ($m_{\tilde G}\mincir 50$ eV) so as to be
cosmologically irrelevant (unless entropy production with a sufficiently
low-reheat temperature dilutes the gravitino abundances). In this case,
however, one would lose the LSP candidate for implementing a CDM--dominated
scenario.

As a concluding remark, we should point out that, from the point of
view of the particle physics model building, we still lack an
exhaustive construction of realistic GMSB schemes, in particular as
far as the details of the messenger sector are concerned. In this
respect we hope that our analysis may constitute a useful guideline
for the intense work which is going on in the GMSB option.

\section*{Acknowledgments} SB wishes to acknowledge SISSA for his
hospitality during several phases of preparation of this work.
MY also acknowledges the hospitality of SISSA where he stayed at a final stage
of this work. The work of MY was partially supported by a grant from Deutsche 
Forschungsgemeinschaft SFB-375-95, the European Commission programs
ERBFMRX-CT96-0045 and CT96-0090, and the Grant-in-Aid for Scientific Research
from the Ministry of Education, Science and Culture of Japan No.09640333.

\newpage
\section*{Appendix}

In this appendix, we summarize the decay widths to gravitino which are
needed  in the calculation of $g_*$ in Section 2.2.  We denote the gluino by
$\tilde g$ (with mass $M_3$), the winos ($U(2)_L$ gauginos) by $\tilde
W^{\pm}$, $\tilde W^0$ (with mass $M_2$), and the bino ($U(1)_Y$ gaugino)
by $\tilde B$ (with mass $M_1$). We ignored possible mixing between the
gauginos and higgsinos.

The decay widths involving the gauginos are
\begin{eqnarray}
\Gamma(\tilde g \rightarrow g +\tilde G)
&=&\frac{1}{48 \pi}\frac{M_3^5}{m_{\tilde G}^2 M_{Pl}^2} \\
\Gamma(\tilde W^{\pm} \rightarrow W^{\pm} +\tilde G)
&=&\frac{1}{48 \pi}\frac{M_2^5}{m_{\tilde G}^2 M_{Pl}^2}
   \left(1-\frac{m_W^2}{M_2^2}\right)^4 ~~~(M_2>m_W) \\
\Gamma(W^{\pm} \rightarrow \tilde W^{\pm}+\tilde G)
&=&\frac{1}{72 \pi}\frac{m_W^5}{m_{\tilde G}^2 M_{Pl}^2}
  \left(1-\frac{M_2^2}{m_W^2}\right)^4 ~~~(M_2<m_W) \\
\Gamma(\tilde B \rightarrow \gamma +\tilde G)
&=&\frac{\cos^2 \theta_W}{48 \pi}\frac{M_1^5}{m_{\tilde G}^2 M_{Pl}^2} \\
\Gamma(\tilde W^0 \rightarrow \gamma +\tilde G)
&=&\frac{\sin^2 \theta_W}{48 \pi}\frac{M_2^5}{m_{\tilde G}^2 M_{Pl}^2} \\
\Gamma(\tilde B \rightarrow Z +\tilde G)
&=&\frac{\sin^2 \theta_W}{48 \pi}\frac{M_1^5}{m_{\tilde G}^2 M_{Pl}^2}
   \left(1-\frac{m_Z^2}{M_1^2}\right)^4 ~~~(M_1>m_Z) \\
\Gamma(Z \rightarrow \tilde B+\tilde G)
&=&\frac{\sin^2 \theta_W}{72 \pi}\frac{m_Z^5}{m_{\tilde G}^2 M_{Pl}^2}
  \left(1-\frac{M_1^2}{m_Z^2}\right)^4 ~~~(M_1<m_Z) \\
\Gamma(\tilde W^0 \rightarrow Z +\tilde G)
&=&\frac{\cos^2 \theta_W}{48 \pi}\frac{M_2^5}{m_{\tilde G}^2 M_{Pl}^2}
   \left(1-\frac{m_Z^2}{M_2^2}\right)^4 ~~~(M_2>m_Z) \\
\Gamma(Z \rightarrow \tilde W^0+\tilde G)
&=&\frac{\cos^2 \theta_W}{72 \pi}\frac{m_Z^5}{m_{\tilde G}^2 M_{Pl}^2}
  \left(1-\frac{M_2^2}{m_Z^2}\right)^4 ~~~(M_2<m_Z), 
\end{eqnarray}
where $m_Z$, $m_W$ are $Z$-- and $W$--gauge boson masses, respectively, and 
$\theta_W$ represents the electro-weak mixing angle. 

The decay width of a slepton with mass $m_{\tilde l}$ to gravitino is given as
\begin{equation} 
\Gamma(\tilde l \rightarrow l +\tilde G)=
\frac{1}{48 \pi} \frac{m_{\tilde l}^5}{m_{\tilde G}^2 M_{Pl}^2}.
\end{equation}
A similar expression is obtained for the decay width of a squark.

\newpage

\section*{Figure captions}
\noindent
{\bf Figure 1.} The shape of the trasfer functions for WDM gravitino
models. Left panel: the effect of varying $g_*$ for $\Omega_0=1$ and
$h=0.5$; solid, dotted and dashed curves correspond to the CDM case,
to $g_*=200$ and $g_*=100$, respectively. Right panel: the effect of
varying the Friedmann background; heavy and light curves corespond to
the CDM and WDM with $g_*=200$ cases, respectively.

\noindent
{\bf Figure 2.} The shape of the transfer function for the warm + hot
DM models. Left panel: the effect of varying $g_*$ at a fixed value of
$\Omega_{\nu}=0.25$. Right panel: the effect of varying $\Omega_\nu$ at a
fixed $g_*=200$.

\noindent
{\bf Figure 3.} The mass--scale dependence of the r.m.s. density
fluctuations within a top--hat sphere. Left and right panels are for
the same models as reported in Figure 1. Heavy and light curves are
for WDM and CDM cases. As for the WDM curves, the value of $M$ at
which they become lighter corresponds to the value of the
free--streaming mass, defined according to eq.(\ref{eq:mfs}).

\noindent
{\bf Figure 4.} Observational constraints for {\sl COBE}--normalized
WDM models, with $g_*=150$, on the $(\Omega_0,h)$ parameter space, for
flat low--density ($\Lambda$WDM) and open (OWDM) models. The 
finely shaded area corresponds to the 95\% c.l.  region allowed by the
cluster abundance, as estimated by Viana \& Liddle \cite{sig8Om} (see
text). The heavy solid curve delimiting the coarsely shaded
area represents the limit of the region allowed by the data about the
$\Omega_{HI}$ in DLAS at $z=4.25$, as given by Storrie--Lombardi et
al.  \cite{Storrie95} (see text); models lying below such curves are
excluded. Horizontal dashed curves connect models having the same age
of the Universe: $t_0=9,11,13,15,17$ Gyrs from upper to lower curves.

\noindent
{\bf Figuree 5.} Observational constraints for {\sl COBE}--normalized
WHDM models, with $g_*=150$, on the $(\Omega_\nu, n_{pr})$ plane, for
$h=0.5$ (left panel) and $h=0.6$ (right panels). A vanishing tensor
mode contribution to CMB temperature anisotropies is assumed for
$n_{pr} < 1$ models. The two shaded areas have the same
meaning as in Figure 4.


\newpage


\begin{thebibliography}{99}

\bibitem{Nilles} For a review, see H.P.~Nilles, {\sl Phys. Rep.} {\bf
    110}, 1 (1984).

\bibitem{SUSYDM-review}
G. Jungman, M. Kamionkowski and K. Griest,
 {\sl Phys. Rep.} {\bf 267}, 195 (1996) and references therein.

\bibitem{Dineetal}
M.~Dine, A.~Nelson, Y.~Nir and Y.~Shirman, {\sl Phys. Rev.} {\bf D53}, 2658 
(1996). \\ 
M.~Dine and A.E.~Nelson,  {\sl Phys. Rev.} {\bf D48}, 1277 (1993). \\
M.~Dine, A.E.~Nelson and Y.~Shirman, {\sl Phys. Rev.} {\bf D51}, 1362 (1995).

\bibitem{cdm} H.~Pagels and J.R.~Primack, {\sl Phys. Rev. Lett.} {\bf
    48}, 233 (1982). \\
    G.R. Blumenthal, S.M. Faber, J.R. Primack and M.J. Rees, {\sl
    Nature} {\bf 311}, 517 (1984).

\bibitem{PeacDodds} J.A. Peacock and S.J. Dodds,
  Mon. Not. R. Astron. Soc. {\bf 267}, 1020 (1994).

\bibitem{cobe} K.M. G\'orski, A.J. Banday, C.L. Bennett, G. Hinshaw,
  A. Kogut, G.F. Smoot, E.L. Wright, {\sl ApJ} {\bf 464}, L11 (1996).

\bibitem{clabu} S.D.M. White, G. Efstathiou and C.S Frenk,
  {\sl Mon. Not. R. Astr. Soc.} {\bf 262}, 1023 (1993).

\bibitem{primack96} J.R. Primack, in {\em Proceeding of the Princeton
    20th Century Annyversary Conference, Critical Dialogues in
    Cosmology},  edited by N. Turok (World Scientific, 1997)

\bibitem{lowom} A.R. Liddle, D.H. Lyth, D. Roberts and P.T.P. Viana,
  {\sl Mon. Not. R. Astr. Soc.} {\bf 278}, 644 (1996). \\ A.R. Liddle,
  D.H. Lyth, P.T.P. Viana and M. White, {\sl Mon. Not. R. Astr. Soc.}
  {\bf 282}, 281 (1996). 
  
\bibitem{BMY} S. Borgani, A. Masiero and M. Yamaguchi, {\sl Phys.
    Lett.} {\bf B386}, 189 (1996). 

\bibitem{KSY} M. Kawasaki, N. Sugiyama and T.
  Yanagida, {\sl Mod. Phys. Lett.} {\bf A12}, 1275 (1997).

\bibitem{colombi} S. Colombi, S. Dodelson and L.M. Widrow,  {\sl ApJ}
{\bf 458}, 1 (1996). 

\bibitem{gauge-80s} M.~Dine, W.~Fischler and M.~Srednicki, {\sl Nucl.
    Phys.} {\bf B189}, 575 (1981).\\
  S.~Dimopoulos and S.~Raby, {\sl Nucl. Phys.} {\bf B192}, 353 (1981). \\
  M.~Dine and W.~Fischler, {\sl Phys. Lett.} {\bf B110}, 227 (1982). \\
  M.~Dine and M.~Srednicki, {\sl Nucl. Phys.} {\bf B202}, 238 (1982). \\
  L.~Alvarez-Gaume, M.~Claudson and M.~Wise, {\sl Nucl. Phys.} {\bf
    B207}, 96 (1982). \\
  C.~Nappi and B.~Ovrut, {\sl Phys. Lett.} {\bf B113}, 175 (1982).

\bibitem{deGMM}
A.~de~Gouv\^{e}a, T.~Moroi and H.~Murayama,
preprint hep-ph/9701244. 

\bibitem{supertrace}
S.~Ferrara, L.~Girardello and F.~Palumbo,
{\sl Phys. Rev.} {\bf D20}, 403 (1979). 

\bibitem{INTY}
K.-I.~Izawa, Y.~Nomura, K.~Tobe and T.~Yanagida, 
preprint hep-ph/9705228. \\
Y. Nomura and K. Tobe, preprint hep-ph/9708377.

\bibitem{no-scale}
E.~Cremmer, S.~Ferrara, C.~Kounnas and D.V.~Nanopoulos,
{\sl Phys. Lett.} {\bf B133}, 61 (1983). \\
J.~Ellis, C.~Kounnas and D.V.~Nanopoulos, 
{\sl Nucl. Phys.} {\bf B241}, 406 (1984); {\sl Nucl. Phys.} 
{\bf B247}, 373 (1984). \\
J.~Ellis, A.~Lahanas, D.V.~Nanopoulos and K.~Tamvakis, {\sl Phys. Lett.} 
{\bf B134}, 439 (1984). \\
J.~Ellis, K.~Enqvist and D.V.~Nanopoulos, {\sl Phys. Lett.} {\bf B147}, 99
(1984).

\bibitem{Fayet}
P.~Fayet, {\sl Phys. Lett.} {\bf 84B}, 421 (1979); 
{\sl Phys. Lett.} {\bf B175}, 471 (1986).



\bibitem{gravitino-problem} 
J. Ellis, J.E. Kim and D.V. Nanopoulos,
{\sl Phys. Lett.} {\bf B145}, 181 (1984). \\ 
R.~Juszkiewicz, J.~Silk
and Z.~Stebbins, {\sl Phys. Lett.} {\bf B158}, 463 (1985). \\
J.~Ellis, D.V.~Nanopoulos and S.~Sarkar, {\sl Nucl. Phys.} {\bf B259},
175 (1985).  
\\ V.S.~Berezinsky, {\sl Phys. Lett.} {\bf B261}, 71
(1991). 
\\ M. Kawasaki and T. Moroi, {\sl Prog. Theor. Phys.} {\bf
93}, 879 (1995).


\bibitem{MMY}
T.~Moroi, H.~Murayama and M.~Yamaguchi, {\sl Phys. Lett.} {\bf B303}, 289
 (1993).

\bibitem{KolbTurner}
E.W. Kolb and M.S. Turner, {\em The Early Universe} (Addison Wesley,
Chicago IL, 1990).

\bibitem{PagelsPrimack}
H.~Pagels and J.R.~Primack, {\sl Phys. Rev. Lett.} {\bf 48}, 233 (1982).

\bibitem{EWbaryogenesis}
M.~Carena, M.~Quiros, A.~Riotto, I.~Vilja and C.E.M.~Wagner, preprint 
hep-ph/9702409. 

\bibitem{CDFevent}
S. Park, "Search for New Phenomena in CDF", 10th Topical Workshop on 
Proton-Antiproton Collider Physics, edited by R. Raja and J. Yoh (AIP Press,
1995). 

\bibitem{gravitino-interpretation}
S. Dimopoulos, M. Dine, S. Raby and S. Thomas, {\sl Phys. Rev. Lett.}
{\bf 76}, 3494 (1996). \\
S. Ambrosanio, G.L. Kane, G.D.~Kribs, S.P.~Martin and S.~Mrenna,
{\sl Phys. Rev. Lett.} {\bf 76}, 3498 (1996); {\sl Phys. Rev.} {\bf D54}, 
5395 (1996). \\
S. Dimopoulos, S. Thomas and J.D. Wells, {\sl Phys. Rev.} {\bf D54}, 
3283 (1996). \\
K.S. Babu, C. Kolda and F. Wilczek, {\sl Phys. Rev.} {\bf 77}, 3070 (1996). \\
J. Lopez and D.V. Nanopoulos, {\sl Mod. Phys. Lett.} {\bf A11}, 2473 (1996);
{\sl Phys. Rev.} {\bf D55}, 4450 (1997).

\bibitem{wdm}
P.J.E. Peebles, {\sl Astrophys. J.} {\bf 258}, 415 (1982). \\
J.R. Bond, A.S. Szalay and M.S. Turner, {\sl Phys. Rev. Lett.} {\bf 48},
1636 (1982). \\
K.A. Olive and M.S. Turner, {\sl Phys. Rev.} {\bf D25}, 213 (1982).


\bibitem{DGP}
G. Dvali, G.F. Giudice and A. Pomarol, 
{\sl Nucl. Phys.} {\bf B478}, 31 (1996). 
\bibitem{maber95} C.P. Ma and E. Bertschinger, {\sl ApJ} {\bf 455}, 7 (1995).
\bibitem{Lidlyth93} A.R. Liddle  and D.H. Lyth, {\sl Phys. Rep.} {\bf 231}, 1
(1993),
\bibitem{numrec} W. Press and S. Teukolsky, ``Numerical Recipes in Fortran''
   (1992), p. 140
\bibitem{bardeen86} J.M. Bardeen, J.R. Bond, N. Kaiser and
  A.S. Szalay, {\sl Astrophys. J.} {\bf 304}, 15 (1986).
\bibitem{sugiyama95} N. Sugiyama, {\sl Astrophys. J. Suppl.} {\bf 100}, 281
  (1995). 


\bibitem{pogstar95} D. Yu Pogosyan and A.A. Starobinski,
    Astrophys. J. {\bf 447}, 465 (1995).
\bibitem{BunnWhite} E.F. Bunn and M. White, Astrophys. J. {\bf 480}, 6
  (1997).
\bibitem{vbulk} A. Dekel, Ann. Rev. Astr. Ap. {\bf 32}, 371 (1994) \\
  M.A. Strauss and J.A. Willick, Phys. Rep. {\bf 261}, 271 (1995) \\
  L.N. da Costa, S. Borgani, W. Freudling, R. Giovanelli, M.P. Haynes,
  J. Salzer and G. Wegner, in preparation (1997).
\bibitem{Wolfe93} A. Wolfe, in {\em Relativistic Astrophysics and
    Particle Cosmology}, edited by C.W. Ackerlof and M.A. Srednicki
  (New York Acad. Sci, New York, 1993).
\bibitem{BartLoeb96} M. Bartelman and A. Loeb, Astrophys. J. {\bf 457},
  529 (1996).
\bibitem{FallPei96} S.M. Fall and Y.C. Pei, in {\em QSO Absorbtion
    Lines} (Springer Verlag, Berlin, 1996)
\bibitem{Storrie95} L.J. Storrie--Lombardi, R.G. McMahon, M.J. Irwin
  and C. Hazard, in {\em Proceeding of the ESO Workshop on QSO
    Absorbtion Lines}, preprint astro--ph/9503089 (1995).
\bibitem{dlasth} H.J. Mo and J. Miralda--Escud\'e,
  Astrophys. J. Lett. {\bf 430}, L25 (1994). \\ G. Kauffmann and
  S. Charlot, Astrophys. J. Lett. {\bf 430}, L97 (1994). \\ C.P. Ma
  and E. Bertschinger, Astrophys. J. Lett. {\bf 434}, L5 (1994). \\
  A. Klypin, S. Borgani, J. Holtzman and J.R. Primack,
  Astrophys. J. {\bf 444}, 1 (1995). \\ S. Borgani, F. Lucchin,
  S. Matarrese and L. Moscardini, Mon. Not. R. Astr. Soc. {\bf 280},
  749 (1996). \\ J.P. Gardner, N. Katz, L. Hernquist and
  D.H. Weinberg, preprint astro--ph/9609072 (1996). \\ M.G. Haenhelt,
  M. Steinmetz and M. Rauch, preprint astro--ph/9706201 (1997).

\bibitem{Proch97} J.X. Prochaska and A.M. Wolfe, preprint
  astro--ph/9704196 (1997)
\bibitem{LaceyCole} C. Lacey and S. Cole, Mon. Not. R. Astr. Soc. {\bf
    271}, 676 (1994).
\bibitem{KitaSuto} T. Kitayama and Y. Suto, Astrophys. J. {\bf 469},
  480 (1996).
\bibitem{PrSc} W.H. Press and P.L. Schechter, Astrophys. J. {\bf 187},
  425 (1974).
\bibitem{sig8Om} P.T.P. Viana and A.R. Liddle,
  Mon. Not. R. Ast. Soc. {\bf 281}, 323 (1996). \\ V.R. Eke, S. Cole
  and C.S. Frenk, Mon. Not. R. Astr. Soc. {\bf 282}, 263 (1996). \\
  U.L. Pen, preprint astro--ph/9610147 (1996).
\bibitem{Freed96} W.L. Freedman, in {\em Proceeding of the Princeton
    20th Century Annyversary Conference, Critical Dialogues in
    Cosmology},  edited by N. Turok (World Scientific, 1997)
\bibitem{globcl} R.G. Gratton, F. Fusi--Pecci, E. Carretta,
  G. Clementini, C. Corsi and M. Lattanzi, preprint astro--ph/9704150
  (1997).
\bibitem{sneutrino} S. Dimopoulos, G.F. Giudice and A. Pomarol, Phys. Lett. 
{\bf B389}, 37 (1996). \\
T. Han and R. Hempfling, preprint hep-ph/9708264.
\bibitem{texture} U.L. Pen, U. Seljak and N. Turok, preprint
  astro--ph/9704165 (1997).
\bibitem{notex} R. Cen, preprint astro--ph/9707240

\end{thebibliography}
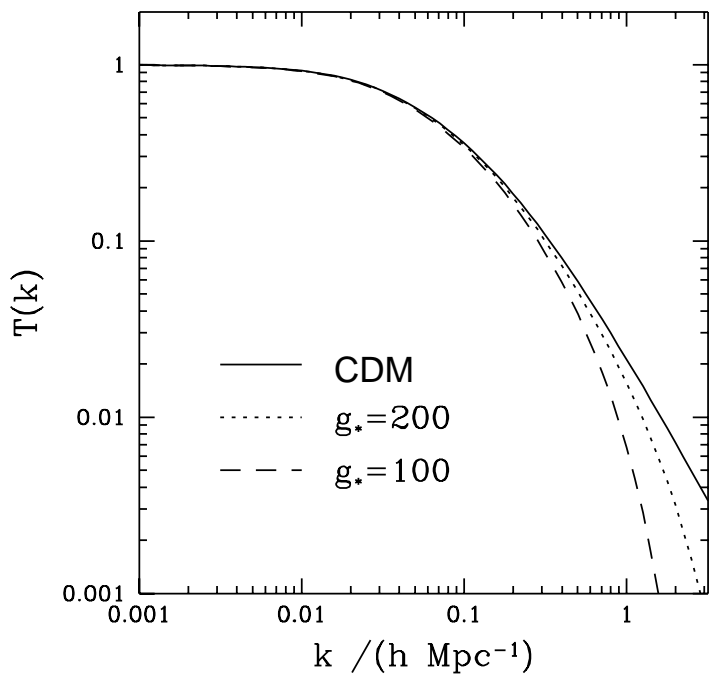
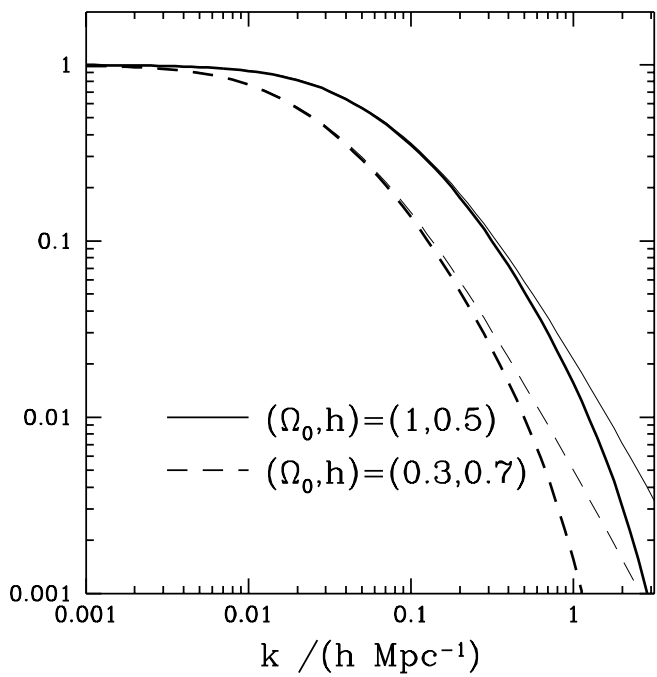
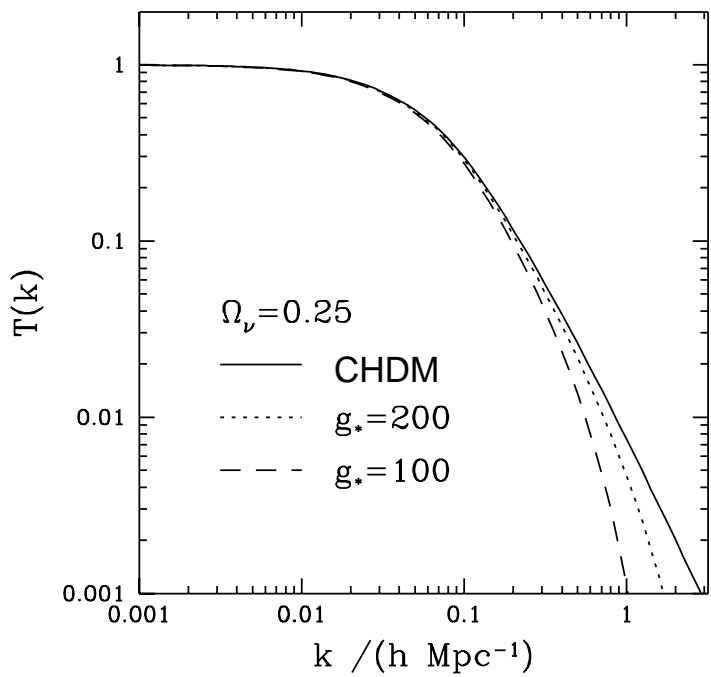
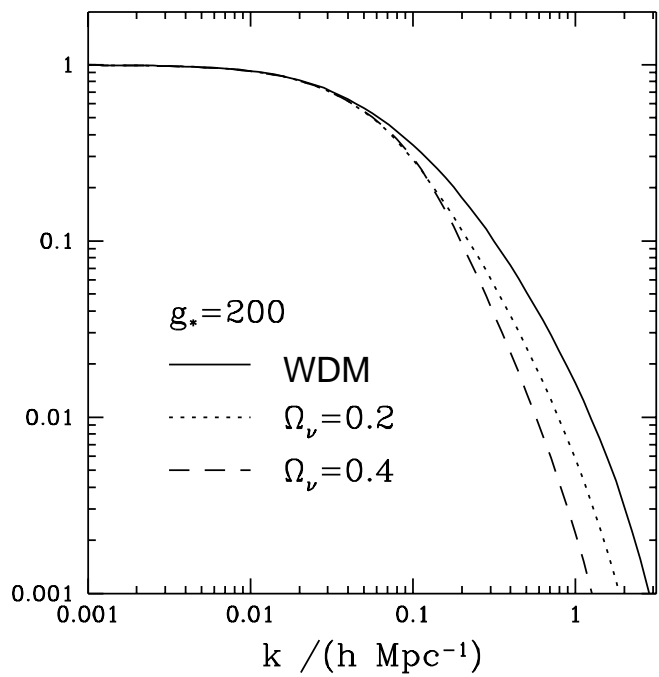
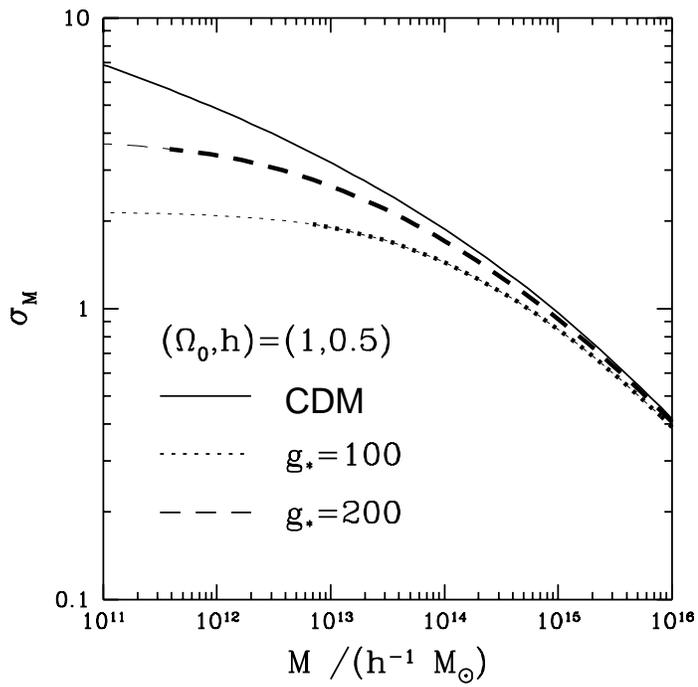
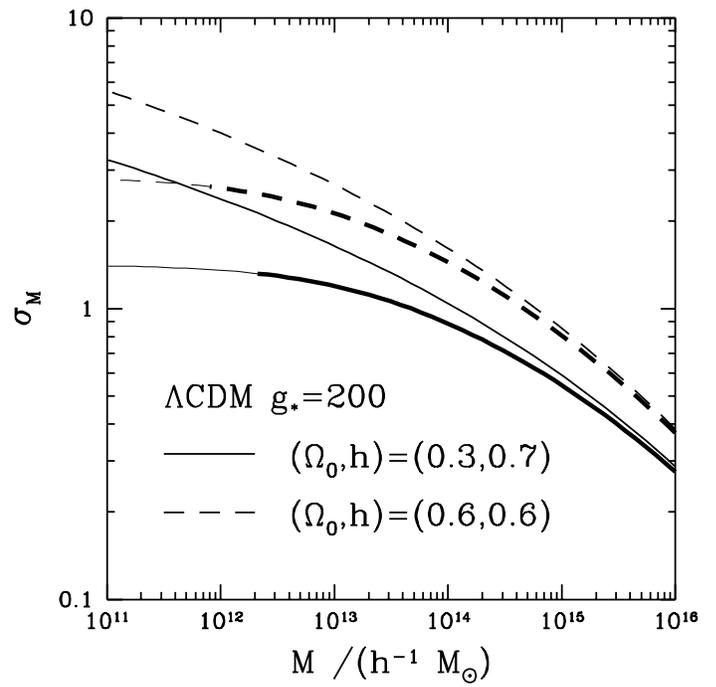
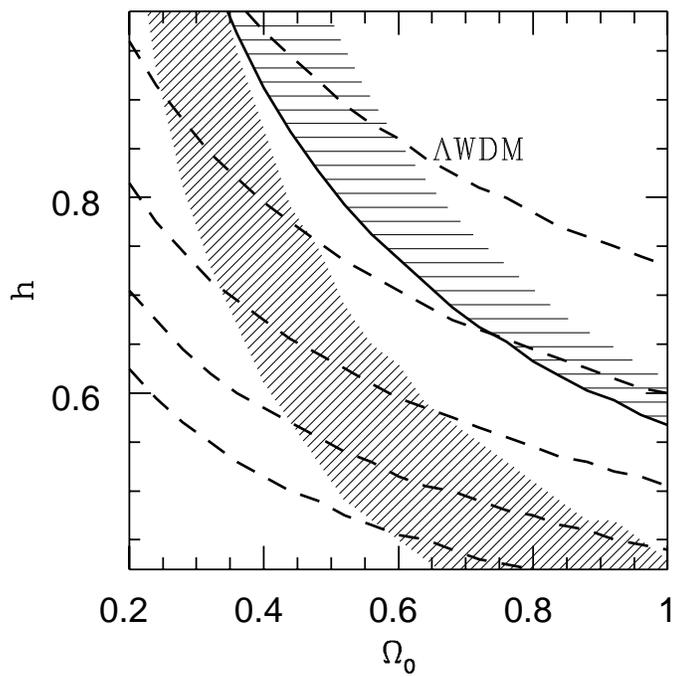
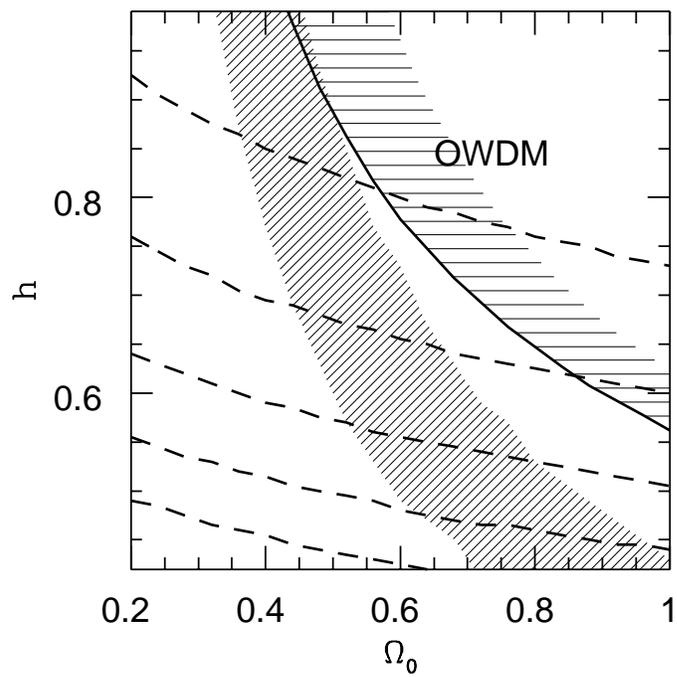
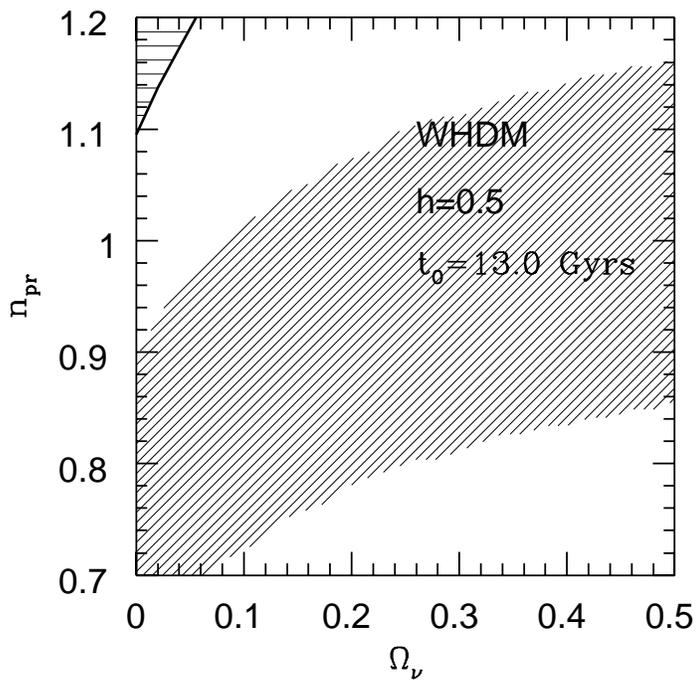
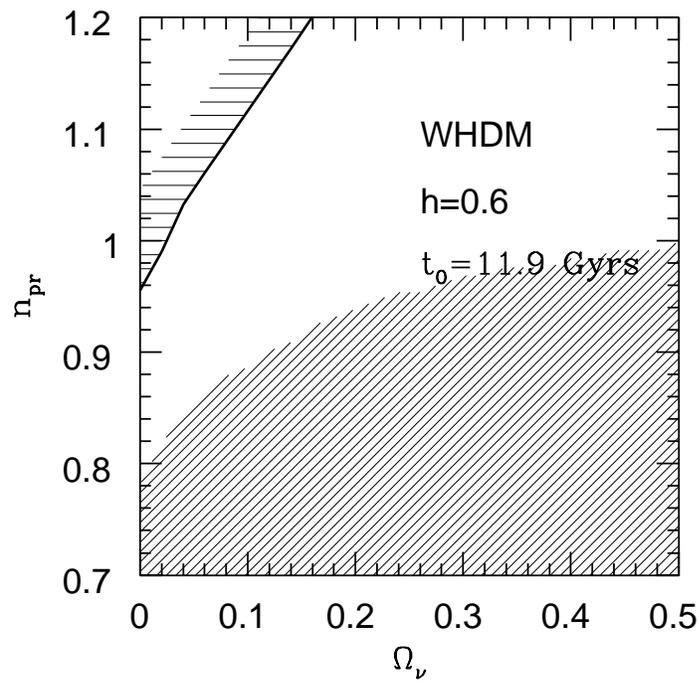
\end{document}